\documentclass[preprint]{aastex63} 
\usepackage{textcomp}  %% for reference author's name
\usepackage{comment}
\usepackage{subfigure}
\submitjournal{ApJ}

\shorttitle{GWneutrioKam2020}
\shortauthors{KamLAND collaboration}

\graphicspath{{./}{figures/}}

\begin{document}

\title{Search for Low-energy Electron Antineutrinos in KamLAND Associated with Gravitational Wave Events}

\correspondingauthor{S.~Obara}
\email{shuhei.obara.d4@tohoku.ac.jp}

%%%%%%[authors]%%%%%%%%%%%%%%%%%%%%%%%%%%%%%%%%%%%%%%%%%%%%%%%
% All university affiliations addresses go here:
\newcommand{\tohoku}{\affiliation{Research Center for Neutrino Science, Tohoku University, Sendai 980-8578, Japan}}
\newcommand{\fris}{\affiliation{Frontier Research Institute for Interdisciplinary Sciences, Tohoku University, Sendai, 980-8578, Japan}}
\newcommand{\gppu}{\affiliation{Graduate Program on Physics for the Universe, Tohoku University, Sendai 980-8578, Japan}}
\newcommand{\ipmu}{\affiliation{Institute for the Physics and Mathematics  of the Universe, The University of Tokyo, Kashiwa 277-8568, Japan}}
\newcommand{\osakarcnp}{\affiliation{Graduate School of Science, Osaka University, Toyonaka, Osaka 560-0043, Japan}}   
\newcommand{\osaka}{\affiliation{Research Center for Nuclear Physics (RCNP), Osaka University, Ibaraki, Osaka 567-0047, Japan}}
\newcommand{\tokushima}{\affiliation{Graduate School of Advanced Technology and Science, Tokushima University, Tokushima, 770-8506, Japan}}
\newcommand{\kyoto}{\affiliation{Department of Physics, Kyoto University, Kyoto 606-8502, Japan}}
\newcommand{\lbl}{\affiliation{Nuclear Science Division, Lawrence Berkeley National Laboratory, Berkeley, CA 94720, USA}}
\newcommand{\hawaii}{\affiliation{Department of Physics and Astronomy, University of Hawaii at Manoa, Honolulu, Hawaii 96822, USA}}
\newcommand{\mituniv}{\affiliation{Massachusetts Institute of Technology, Cambridge, Massachusetts 02139, USA}}
\newcommand{\bu}{\affiliation{Boston University, Boston, Massachusetts 02215, USA}}
\newcommand{\tennessee}{\affiliation{Department of Physics and Astronomy,  University of Tennessee, Knoxville, Tennessee 37996, USA}}
\newcommand{\tunl}{\affiliation{Triangle Universities Nuclear Laboratory,  Durham, North Carolina 27708, USA and Physics Departments at Duke University, North Carolina Central University, and the University of North Carolina at Chapel Hill}}    
\newcommand{\chapehill}{\affiliation{The University of North Carolina at Chapel Hill, Chapel Hill, North Carolina 27599, USA}}

\newcommand{\northcarolina}{\affiliation{North Carolina Central University, Durham, North Carolina 27701, USA}}
\newcommand{\duke}{\affiliation{Physics Department at Duke University, Durham, North Carolina 27705, USA}}
\newcommand{\seattle}{\affiliation{Center for Experimental Nuclear Physics and Astrophysics, University of Washington, Seattle, Washington 98195, USA}}
\newcommand{\nikhef}{\affiliation{Nikhef, Science Park 105, 1098 XG Amsterdam, The Netherlands}}
\newcommand{\virginia}{\affiliation{Center for Neutrino Physics, Virginia Polytechnic Institute and State University, Blacksburg, Virginia 24061, USA}}

\newcommand{\currentKozkov}{\affiliation{Present address: National Research Nuclear University ``MEPhI'' (Moscow Engineering Physics Institute), Moscow, 115409, Russia}}
\newcommand{\currentDima}{\affiliation{Present address: Department of Physics and Astronomy, University of Alabama, Tuscaloosa, Alabama 35487, USA and Institute for Nuclear Research of NASU, 03028 Kyiv, Ukraine}}

\newcommand{\currentHayashida}{\affiliation{Present address: Imperial College London, Department of Physics, Blackett Laboratory, London SW7 2AZ, UK}}
\newcommand{\currentUeshima}{\affiliation{Present address: National Institutes for Quantum and Radiological Science and Technology (QST), Hyogo 679-5148, Japan}}
\newcommand{\currentTakemoto}{\affiliation{Present address: Kamioka Observatory, Institute for Cosmic-Ray Research, The University of Tokyo, Hida, Gifu 506-1205, Japan}}
\newcommand{\currentAobo}{\affiliation{Present Address: UNC Physics and Astronomy, 120 E. Cameron Ave., Phillips Hall CB3255, Chapel Hill, NC 27599}}

%%
%%% Japan
%%Tohoku 
\author{S.~Abe}\tohoku
\author{S.~Asami}\tohoku
\author{A.~Gando}\tohoku
\author{Y.~Gando}\tohoku
\author{T.~Gima}\tohoku 
\author{A.~Goto} \tohoku
\author{T.~Hachiya}\tohoku
\author{K.~Hata} \tohoku
\author{S.~Hayashida}\tohoku \currentHayashida
\author{K.~Hosokawa} \tohoku
\author{K.~Ichimura} \tohoku  
\author{S.~Ieki} \tohoku
\author{H.~Ikeda}\tohoku
\author{K.~Inoue}\tohoku \ipmu 
%\author[0000-0001-9271-2301]{K.~Ishidoshiro}\tohoku
\author{K.~Ishidoshiro}\tohoku
\author{Y.~Kamei} \tohoku
%\author[0000-0003-2350-2786]{N.~Kawada} \tohoku
\author{N.~Kawada} \tohoku
\author{Y.~Kishimoto} \tohoku \ipmu
\author{T.~Kinoshita} \tohoku 
\author{M.~Koga}\tohoku \ipmu 
\author{N.~Maemura}\tohoku
\author{T.~Mitsui}\tohoku
\author{H.~Miyake}\tohoku
\author{K.~Nakamura}\tohoku % Kengo Nakamura
\author{K.~Nakamura}\tohoku % Kosuke Nakamura
\author{R.~Nakamura}\tohoku
\author{H.~Ozaki}\tohoku \gppu
\author{T.~Sakai} \tohoku 
\author{H.~Sambonsugi}\tohoku
\author{I.~Shimizu}\tohoku
\author{J.~Shirai}\tohoku
\author{K.~Shiraishi}\tohoku
\author{A.~Suzuki}\tohoku
\author{Y.~Suzuki}\tohoku %Suzuki Yuya
\author{A.~Takeuchi}\tohoku
\author{K.~Tamae}\tohoku
\author{K.~Ueshima}\tohoku \currentUeshima 
\author{Y.~Wada}\tohoku
\author{H.~Watanabe}\tohoku
\author{Y.~Yoshida} \tohoku
%%FRIS(tohoku)
%\author[0000-0003-3488-3553]{S.~Obara}\fris %% corresponding author
\author{S.~Obara}\fris %% corresponding author
%%IPMU 
\author{A.~Kozlov}\ipmu \currentKozkov
\author{D.~Chernyak}\ipmu \currentDima
%%Osaka 
\author{Y.~Takemoto}\osakarcnp \currentTakemoto
\author{S.~Yoshida}\osakarcnp
\author{S.~Umehara}\osaka
%%Tokushima 
\author{K.~Fushimi}\tokushima
%%Kyoto 
\author{A.K.~Ichikawa}\kyoto
\author{K.Z.~Nakamura}\kyoto
\author{M.~Yoshida}\kyoto 
%%
%%
%%%Non-Japan
%%
%%Berkeley 
\author{B.~E.~Berger}\lbl \ipmu
\author{B.K.~Fujikawa}\lbl \ipmu
%%Hawaii 
\author{J.G.~Learned}\hawaii
\author{J.~Maricic}\hawaii
%%MIT 
\author{S.~Axani}\mituniv
\author{L.~A.~Winslow}\mituniv
\author{Z.~Fu}\mituniv
\author{J.~Ouellet}\mituniv
%% UT 
\author{Y.~Efremenko}\tennessee \ipmu
%% TUNL 
\author{H.~J.~Karwowski}\tunl \chapehill
\author{D.~M.~Markoff}\tunl \northcarolina
\author{W.~Tornow}\tunl \duke \ipmu
%\author[0000-0002-4844-9339]{A.~Li}\chapehill %\bu \currentAobo
\author{A.~Li}\chapehill %\bu \currentAobo
%% WT 
\author{J.~A.~Detwiler}\seattle \ipmu
\author{S.~Enomoto}\seattle \ipmu
%% Nikef 
\author{M.P.~Decowski}\nikhef \ipmu
%% Boston 
\author{C.~Grant}\bu
%% Virginia 
\author{T.~O'Donnell}\virginia
\author{S.~Dell'Oro}\virginia

\collaboration{99}{(KamLAND Collaboration)}

%%%%%[abst]%%%%%%%%%%%%%%%%%%%%%%%%%%%%%%%%%%%%%%%%%%%%%%%%%%
\begin{abstract}
We present the results of a search for MeV-scale electron antineutrino events in KamLAND in coincident with the 60~gravitational wave events/candidates reported by the LIGO/Virgo collaboration during their second and third observing runs. 
We find no significant coincident signals within a $\pm$500\,s timing window from each gravitational wave
and present 90\% C.L. upper limits on the electron antineutrino fluence between $10^{8}$--$10^{13}\,{\rm cm}^2$ for neutrino energies in the energy range of 1.8--111\,MeV.
\end{abstract}

%%%%%[keywords]%%%%%%%%%%%%%%%%%%%%%%%%%%%%%%%%%%%%%%%%%%%%%%%%%%

%%%%%%%%%%%%%%%%%%%%%%%%%%%%%%%%%%%%%%%%%%%%%%%%%%%%%%%
\section{Introduction} \label{sec:intro}

In 2015, gravitational waves (GWs) were first detected by the Advanced Laser Interferometer Gravitational-wave Observatory (LIGO)~\citep{abbott2016}. 
This event was shown to have originated from the merger of a binary black hole (BBH) system.
Nearly two years earlier, the IceCube collaboration published the first observational evidence for high-energy astrophysical neutrinos~\citep{icecube2013evidence}. 
The gravitational and weak forces along with the electromagnetic were added to the astronomical observations, beginning a new era of extra-galactic multi-messenger astronomy.

In 2017, LIGO detected an event consistent with a comparably nearby binary neutron-star (BNS) merger~\citep{abbott2017gw170817}. 
Within seconds of the GW, the electromagnetic counterpart was observed by the \textit{Fermi} Gamma Ray Burst Monitor~\citep{GBM:2017lvd}, making this the first GW multi-messenger event. 
The online neutrino telescopes -- including IceCube, ANTARES, and the Pierre Auger Observatory -- did not detect any directionally coincident high-energy (GeV--EeV) neutrinos or an MeV neutrino burst signal~\citep{ANTARES:2017bia}.  
While no coincident neutrinos were found, this is consistent with model predictions for the merger~\citep{Kimura:2017kan}.  
In contrast with 
BBH mergers, BNS mergers are expected to emit neutrinos at both GeV and MeV energies~\citep{Meszaros:2015krr}. 
MeV-scale neutrinos would be produced by the hot collapsing fireball at the beginning of a gamma-ray burst~\citep{Sahu:2005zh}, so we can be confident that they must be produced when there is collapsing matter outside of a black hole. 
These neutrinos are modeled in several ways, but have energies on the order of the dominant photon energy, and are considerably more numerous than the emitted photons~\citep{Halzen:1996qw}. 
In the case of a post-merger neutron star remnant, thermal emission of MeV neutrinos is expected as the remnant cools~\citep{foucart2016low}. 
The neutron rich environment also suggests a brighter $\bar{\nu}_e$ flux than the $\nu_e$ flux~\citep{PhysRevD.97.103001}.

Recently, the LIGO/Virgo collaboration published their event catalog~\citep{PhysRevX.9.031040}, including the full dataset from their first and second observing runs, LIGO-O1 and LIGO-O2 respectively.
During the third observing run, LIGO-O3, the LIGO/Virgo collaboration initiated the online GW candidate event database (GraceDB)~\citep{GraceDB}, providing public alerts and a centralized location for aggregating and retrieving event information.
For such transient GW events, various neutrino detectors reported correlation searches: Super-Kamiokande~\citep{Abe_2016,Abe_2018}, Borexino~\citep{Agostini_2017}, NOvA~\citep{PhysRevD.101.112006},
Bikal-GVD Neutrino Telescope~\citep{Avrorin2018},
Daya~Bay~\citep{an2020search}, XMASS~\citep{xmass2020gw}, and IceCube/ANTARES~\citep{adrian2016high, ANTARES:2017iky, Aartsen_2020}. 
The Kamioka Liquid scintillator AntiNeutrino Detector (KamLAND) has also performed a search for electron antineutrinos in coincident with gravitational waves GW150914 and GW151226, and then candidate event LVT151012~\citep{Gando_2016}.

In this paper, we present an updated coincidence search for MeV-scale electron antineutrinos in KamLAND associated with the observed GW events in LIGO-O2 (2016 November 30 to 2017 August 25) and LIGO-O3 (2019 April 1 to March 27).

%%%%%%%%%%%%%%%%%%%%%%%%%%%%%%%%%%%%%%%%%%%%%%%%%%%%%%%
\section{KamLAND detector} \label{sec:detector} 

KamLAND is a large volume liquid scintillator neutrino detector located at the Kamioka mine, 1\,km underground from the top of Mt.~Ikenoyama in Gifu Prefecture, Japan.
The KamLAND detector consists of a cylindrical 10\,m radius $\times$ 20\,m height water-Cerenkov outer detector for cosmic-ray muon veto, a 9\,m radius stainless steel spherical tank that mounts 1325\,17-inch and 554\,20-inch photomultiplier-tubes (PMTs), and a 6.5\,m radius Nylon/EVOH outer balloon filled with approximately 1\,kton of ultra-pure liquid scintillator. 
The liquid scintillator is composed of 20\% Pseudocumene (1,2,4-Trimethylbenzene, ${\rm C_{9}H_{12}}$), 80\% Dodecane (N-12, ${\rm C_{12}H_{26}}$), and 1.36\,g/l PPO (2,5-Diphenyloxazole, ${\rm C_{15}H_{11}NO}$). 
Further details of the KamLAND detector are summarized in~\citet{Suzuki2014}.

KamLAND began its data acquisition in 2002 March. 
The detector was upgraded in 2011 August to include a drop-shaped 1.5\,m-radius nylon inner balloon filled with approximately 400\,kg of purified xenon loaded in liquid scintillator~\citep{PhysRevLett.117.082503}. 
In this configuration, known as the KamLAND-Zen~400 experiment, KamLAND searched for neutrinoless double-beta decay until 2015 December, at which point the inner balloon was removed. 
Subsequently in 2018 May, a further upgrade to the KamLAND-Zen~800 experiment ensued with the addition of a 1.9\,m-radius inner balloon, containing approximately 800\,kg of purified xenon.

Electronic boards record the digitized PMT waveforms and provid the corresponding time stamp based on a 40\,MHz internal clock. 
All internal clocks are synchronized to the Unix Time Stamp on every 32nd pulse per second (1\,PPS) trigger from a Global Positioning System receiver, located at the entrance to the Kamioka mine. 
Uncertainties in the absolute trigger time stamp accuracy are less than $\mathcal{O}(100)\,\mu$s, derived from the signal transportation into the mine, optical/electrical signal conversion, and triggering electronics, which is negligibly small for this coincidence search.

The interaction vertex and energy deposition are reconstructed using the measured PMT charge and timing information. 
At low energies, the detector calibrations are performed using various radioactive sources: $^{60}$Co, $^{68}$Ge, $^{203}$Hg, $^{65}$Zn, $^{241}$Am$^{9}$Be, $^{137}$Cs, and $^{210}$Po$^{13}$C. 
At higher energies ($>$10\,MeV), the energy response is calibrated using spallation-produced $^{12}$B/$^{12}$N. 
Daily stability measurements are performed using the 2.2\,MeV gamma ray emitted from a spallation-neutron capture on a proton~\citep{PhysRevC.81.025807}.
The reconstructed energy and interaction vertex resolution are evaluated as $6.4\%/\sqrt{E\,({\rm MeV})}$ and $\sim 12\,{\rm\,cm}/\sqrt{E\,({\rm MeV})}$~\citep{PhysRevD.88.033001}, respectively.

The primary radioactive backgrounds found in the liquid scintillator are $(5.0\pm0.2)\times 10^{-18}$\,g/g ($93\pm4$\,nBq/m$^3$) of $^{238}$U and $(1.8\pm0.1)\times10^{-17}$\,g/g ($59\pm4$\,nBq/m$^3$) of $^{232}$Th~\citep{PhysRevC.92.055808}.

During the period in which LIGO-O2 and LIGO-O3 were collecting data, the KamLAND detector had an average livetime efficiency of $\epsilon_{\rm{live}} = 0.878$. 
For all but one GW event in LIGO-O2, GW170608, the KamLAND detector was actively taking physics data. 
Whereas, three GW events in LIGO-O3 (S191213g, S191215w, and S191216ap) overlapped with
an unusual detector condition period.
Table~\ref{tab:event1} and Table~\ref{tab:event2} summarize the GW events used in this analysis during their respective observing runs, along with the KamLAND detector status.

\begin{deluxetable*}{chcccc}[htbp]
\tablenum{1}
\label{tab:event1}
\tablecaption{The gravitational wave event list for LIGO-O2~\citep{PhysRevX.9.031040} and along with the KamLAND detector status. The three events in which KamLAND has already published the results for a coincidence search~\citep{Gando_2016} are not included in this table.
}
\tablewidth{0pt}
\tablehead{
    \colhead{Gravitational wave} &
    \nocolhead{ref} &
    \colhead{Date and time (UTC)} &
    \colhead{Distance (Mpc)} &
    \colhead{Source}  &
    \colhead{KamLAND status}
}
\startdata
GW170104 & PhysRevX.9.031040 & 2017 January 4, 10:11:58.6 & $990^{+440}_{-430}$    & BBH & running \\
GW170608 & PhysRevX.9.031040 & 2017 June 8, 02:01:16.5 & $320^{+120}_{-110}$    & BBH & unusual data condition \\
GW170729 & PhysRevX.9.031040 & 2017 July 29, 18:56:29.3 & $2840^{+1400}_{-1360}$ & BBH & running \\
GW170809 & PhysRevX.9.031040 & 2017 August 9, 08:28:21.8 & $1030^{+320}_{-390}$   & BBH & running \\
GW170814 & PhysRevX.9.031040 & 2017 August 14, 10:30:43.5 & $600^{+150}_{-220}$    & BBH & running \\
GW170817 & PhysRevX.9.031040 & 2017 August 17, 12:41:04.4 & $40^{+7}_{-15}$        & BNS & running \\
GW170818 & PhysRevX.9.031040 & 2017 August 18, 02:24:09.1 & $1060^{+420}_{-380}$   & BBH & running \\
GW170823 & PhysRevX.9.031040 & 2017 August 23, 13:13:58.5 & $1940^{+970}_{-900}$   & BBH & running \\
\enddata
\end{deluxetable*}

\begin{deluxetable*}{chchc}[htbp]
\tablenum{2}
\label{tab:event2}
\tablecaption{The gravitational wave event list for LIGO-O3~\citep{PhysRevX.9.031040} and KamLAND detector status. Data was extracted from GraceDB~\citep{GraceDB}. 
The retracted events are not shown here.
%KL-Zen~800 indicates the KamLAND-Zen 800 detector configuration.
}
\tablewidth{0pt}
\tabletypesize{\scriptsize}
\tablehead{
    \colhead{Gravitational wave} &
    \nocolhead{refs} &
    \colhead{Date and Time (UTC)} &
    \nocolhead{GW source} &
    \colhead{KamLAND status} 
}
\startdata
S190408an & GraceDB & 2019 April 8, 18:18:02 & BBH ($>$99\%) & running \\ %(KL-Zen 800)\\
S190412m  & GraceDB & 2019 April 12, 05:30:44 & BBH ($>$99\%) & running \\ %(KL-Zen 800)\\
S190421ar & GraceDB & 2019 April 21, 21:38:56 & BBH (97\%), Terrestrial (3\%) & running \\ %(KL-Zen 800)\\
S190425z  & GraceDB & 2019 April 25, 08:18:05 & BNS ($>$99\%) & running \\ %(KL-Zen 800)\\
S190426c  & GraceDB & 2019 April 26, 15:21:55 & \begin{tabular}{c} BNS (49\%), MassGap (24\%),\\ Terrestrial (14\%), NSBH (13\%) \end{tabular} & running \\ %(KL-Zen 800)\\
S190503bf & GraceDB & 2019 May 3, 18:54:04  & BBH (96\%), MassGap (3\%) & running \\ %(KL-Zen 800)\\
S190510g  & GraceDB & 2019 May 10, 02:59:39 & Terrestrial (58\%), BNS (42\%) & running \\ %(KL-Zen 800)\\
S190512at & GraceDB & 2019 May 12, 18:07:14 & BBH (99\%), Terrestrial (1\%) & running \\ %(KL-Zen 800)\\
S190513bm & GraceDB & 2019 May 13, 20:54:28 & BBH (94\%), MassGap (5\%) & running \\ %(KL-Zen 800)\\
S190517h  & GraceDB & 2019 May 17, 05:51:01 & BBH (98\%), MassGap (2\%) & running \\ %(KL-Zen 800)\\
S190519bj & GraceDB & 2019 May 19, 15:35:44 & BBH (96\%), Terrestrial (4\%) & running \\ %(KL-Zen 800)\\
S190521g  & GraceDB & 2019 May 21, 03:02:29 & BBH (97\%), Terrestrial (3\%) & running \\ %(KL-Zen 800)\\
S190521r  & GraceDB & 2019 May 21, 07:43:59 & BBH ($>$99\%) & running \\ %(KL-Zen 800)\\
S190602aq & GraceDB & 2019 June 2, 17:59:27 & BBH (99\%) & running \\ %(KL-Zen 800)\\
S190630ag & GraceDB & 2019 June 30, 18:52:05 & BBH (94\%), MassGap (5\%) & running \\ %(KL-Zen 800)\\
S190701ah & GraceDB & 2019 July 1, 20:33:06  & BBH (93\%), Terrestrial (7\%) & running \\ %(KL-Zen 800)\\
S190706ai & GraceDB & 2019 July 6, 22:26:41 & BBH (99\%), Terrestrial (1\%) & running \\ %(KL-Zen 800)\\
S190707q  & GraceDB & 2019 July 7, 09:33:26 & BBH ($>$99\%) & running \\ %(KL-Zen 800)\\
S190718y  & GraceDB & 2019 July 18, 14:35:12  & Terrestrial (98\%), BNS (2\%) & running \\ %(KL-Zen 800)\\
S190720a  & GraceDB & 2019 July 20, 00:08:36 & BBH (99\%), Terrestrial (1\%) & running \\ %(KL-Zen 800)\\
S190727h  & GraceDB & 2019 July 27, 06:03:33  & BBH (92\%), Terrestrial (5\%), MassGap (3\%) & running \\ %(KL-Zen 800)\\
S190728q  & GraceDB & 2019 July 28, 06:45:10 & BBH (95\%), MassGap (5\%) & running \\ %(KL-Zen 800)\\
S190814bv & GraceDB & 2019 August 14, 21:10:39 & NSBH ($>$99\%) & running \\ %(KL-Zen 800)\\
S190828j  & GraceDB & 2019 August 28, 06:34:05 & BBH ($>$99\%) & running \\ %(KL-Zen 800)\\
S190828l  & GraceDB & 2019 August 28, 06:55:09 & BBH ($>$99\%) & running \\ %(KL-Zen 800)\\
S190901ap & GraceDB & 2019 September 1, 23:31:01 & BNS (86\%), Terrestrial (14\%) & running \\ %(KL-Zen 800)\\
S190910d  & GraceDB & 2019 September 10, 01:26:19  & NSBH (98\%), Terrestrial (2\%) & running \\ %(KL-Zen 800)\\
S190910h  & GraceDB & 2019 September 10, 08:29:58  & BNS (61\%), Terrestrial (39\%) & running \\ %(KL-Zen 800)\\
S190915ak & GraceDB & 2019 September 15, 23:57:02 & BBH (99\%) & running \\ %(KL-Zen 800)\\
S190923y  & GraceDB & 2019 September 23, 12:55:59 & NSBH (68\%), Terrestrial (32\%) & running \\ %(KL-Zen 800)\\
S190924h  & GraceDB & 2019 September 24, 02:18:46 & MassGap ($>$99\%) & running \\ %(KL-Zen 800)\\
S190930s  & GraceDB & 2019 September 30, 13:35:41 & MassGap (95\%), Terrestrial (5\%) & running \\ %(KL-Zen 800)\\
S190930t  & GraceDB & 2019 September 30, 14:34:07 & NSBH (74\%), Terrestrial (26\%) & running \\ %(KL-Zen 800)\\
S191105e  & GraceDB & 2019 November 5, 14:35:21 & BBH (95\%), Terrestrial (5\%) & running \\ %(KL-Zen 800)\\
S191109d  & GraceDB & 2019 November 9, 01:07:17 & BBH ($>$99\%) & running \\ %(KL-Zen 800)\\
S191129u  & GraceDB & 2019 November 29, 13:40:29 & BBH ($>$99\%) & running \\ %(KL-Zen 800)\\
S191204r  & GraceDB & 2019 December 4, 17:15:26 & BBH ($>$99\%) & running \\ %(KL-Zen 800)\\
S191205ah & GraceDB & 2019 December 5, 21:52:08 & NSBH (93\%), Terrestrial (7\%) & running \\ %(KL-Zen 800)\\
S191213g  & GraceDB & 2019 December 13, 15:59:05 & BNS (77\%), Terrestrial (23\%) & unusual data condition \\
S191215w  & GraceDB & 2019 December 15, 22:30:52 & BBH ($>$99\%) & unusual data condition \\
S191216ap & GraceDB & 2019 December 16, 21:33:38 & BBH (99\%) & unusual data condition \\
S191222n  & GraceDB & 2019 December 22, 03:35:37 & BBH ($>$99\%) & running \\ %(KL-Zen 800)\\
S200105ae & GraceDB & 2020 January 5, 16:24:26  & Terrestrial (97\%), NSBH (3\%) & running \\ %(KL-Zen 800)\\
S200112r  & GraceDB & 2020 January 12, 15:58:38 & BBH ($>$99\%) & running \\ %(KL-Zen 800)\\
S200114f  & GraceDB & 2020 January 14, 02:08:18 & -- & running \\ %(KL-Zen 800)\\
S200115j  & GraceDB & 2020 January 15, 04:23:09 & MassGap ($>$99\%) & running \\ %(KL-Zen 800)\\
S200128d  & GraceDB & 2020 January 28, 02:20:11  & BBH (97\%), Terrestrial (3\%) & running \\ %(KL-Zen 800)\\
S200129m  & GraceDB & 2020 January 29, 06:54:58 & BBH ($>$99\%) & running \\ %(KL-Zen 800)\\
S200208q  & GraceDB & 2020 February 8, 13:01:17 & BBH (99\%) & running \\ %(KL-Zen 800)\\
S200213t  & GraceDB & 2020 February 13, 04:10:40 & BNS (63\%), Terrestrial (37\%) & running \\ %(KL-Zen 800)\\
S200219ac & GraceDB & 2020 February 19, 09:44:15 & BBH (96\%), Terrestrial (4\%) & running \\ %(KL-Zen 800)\\
S200224ca & GraceDB & 2020 February 24, 22:22:34  & BBH ($>$99\%) & running \\ %(KL-Zen 800)\\
S200225q  & GraceDB & 2020 February 25, 06:04:21 & BBH (96\%), Terrestrial (4\%) & running \\ %(KL-Zen 800)\\
S200302c  & GraceDB & 2020 March 2, 01:58:11 & BBH (89\%), Terrestrial (11\%) & running \\ %(KL-Zen 800)\\
S200311bg & GraceDB & 2020 March 11, 11:58:53 & BBH ($>$99\%) & running \\ %(KL-Zen 800)\\
S200316bj & GraceDB & 2020 March 16, 21:57:56 & MassGap ($>$99\%) & running \\ %(KL-Zen 800)\\
\enddata
\end{deluxetable*}

%%%%%%%%%%%%%%%%%%%%%%%%%%%%%%%%%%%%%%%%%%%%%%%%%%%%%%%
\section{Electron antineutrino selection and background estimation} \label{sec:antiv}
In this analysis, we focus on KamLAND events induced by the electron antineutrino inverse beta-decay (IBD) reaction ($\bar{\nu}_e + p \rightarrow e^+ + n$) with 1.8\,MeV neutrino energy threshold.
The IBD candidate events can be selected by the delayed coincidence (DC) signature: scintillation light from the positron and its annihilation gamma-rays as a prompt signal, and a 2.2\,MeV (4.9\,MeV) gamma ray from neutron capture on a proton (carbon-12) as a 207.5\,$\pm$\,2.8\,$\mu$s delayed signal~\citep{PhysRevC.81.025807}.
The incident neutrino energy ($E_{\nu}$) is computed from the reconstructed prompt energy ($E_{\rm prompt}$) with energy and momentum conservation in the reaction as 
$E_{\nu} \simeq E_{\rm prompt} + 0.78\,{\rm MeV} + T_n$, where $T_n$ represents the neutron kinetic energy.

The energy range of this analysis is selected to be $E_{\mathrm{prompt}}$ between 0.9--100.0\,MeV, with a delayed neutron capture on a proton (carbon-12) energy between 1.8--2.6\,MeV (4.4--5.6\,MeV). 
Accidental backgrounds are suppressed by imposing a spatial and time correlation between the prompt and delayed signals. 
In particular, the reconstructed vertex and time difference between the prompt and delayed signals must be within 200\,cm and 0.5--1000\,$\mu$s of each other. 
All events must be reconstructed in the fiducial volume region 6 m radius from the center, 
corresponding to a total number of target protons of $N_{T} = (5.98\,\pm\,0.13) \times 10^{31}$.  
Muon and spallation vetoes are applied after the interaction of a cosmic-ray muon, which occur at a rate of approximately 0.34\,Hz in KamLAND.
Further details regarding the event selection can be found in previous KamLAND analyses~\citep{Gando2011, PhysRevD.88.033001,asakura2015study,Gando_2016}.
A likelihood-based signal selection distinguishes electron antineutrino DC pairs from accidental coincidence backgrounds for a few to several MeV energy range.
This has been updated from the previous analyses considering the accidental coincidence event rates, upgraded detector conditions of the outer detector refurbishment~\citep{Ozaki:2016fmr}, inner balloon installation for KamLAND-Zen~800~\citep{GandoY_2020}, and the activity of Japanese nuclear reactors.

From 2018 May onwards (the KamLAND-Zen 800 phase) -- in order to avoid unexpected background contamination due to the xenon-loaded liquid scintillator, inner balloon body, and suspending ropes -- the inner balloon region is vetoed for the delayed event. 
The inner-balloon cut regions are: a 2.5\,m radius spherical volume centered in the detector and a 2.5\,m radius vertical cylindrical volume in the upper-half of detector.
In this analysis, the effect of this additional inner balloon cut is considered as a selection efficiency suppression for the delayed event rather than a change in the number of target protons for the prompt event.  
Therefore, the total selection efficiencies are different between the KamLAND datasets corresponding to the periods operating during LIGO-O2 (without the inner balloon cut) and LIGO-O3 (with the additional inner balloon cut).

\begin{figure}[htbp]
    \centering
    \includegraphics[width=1.0\linewidth]{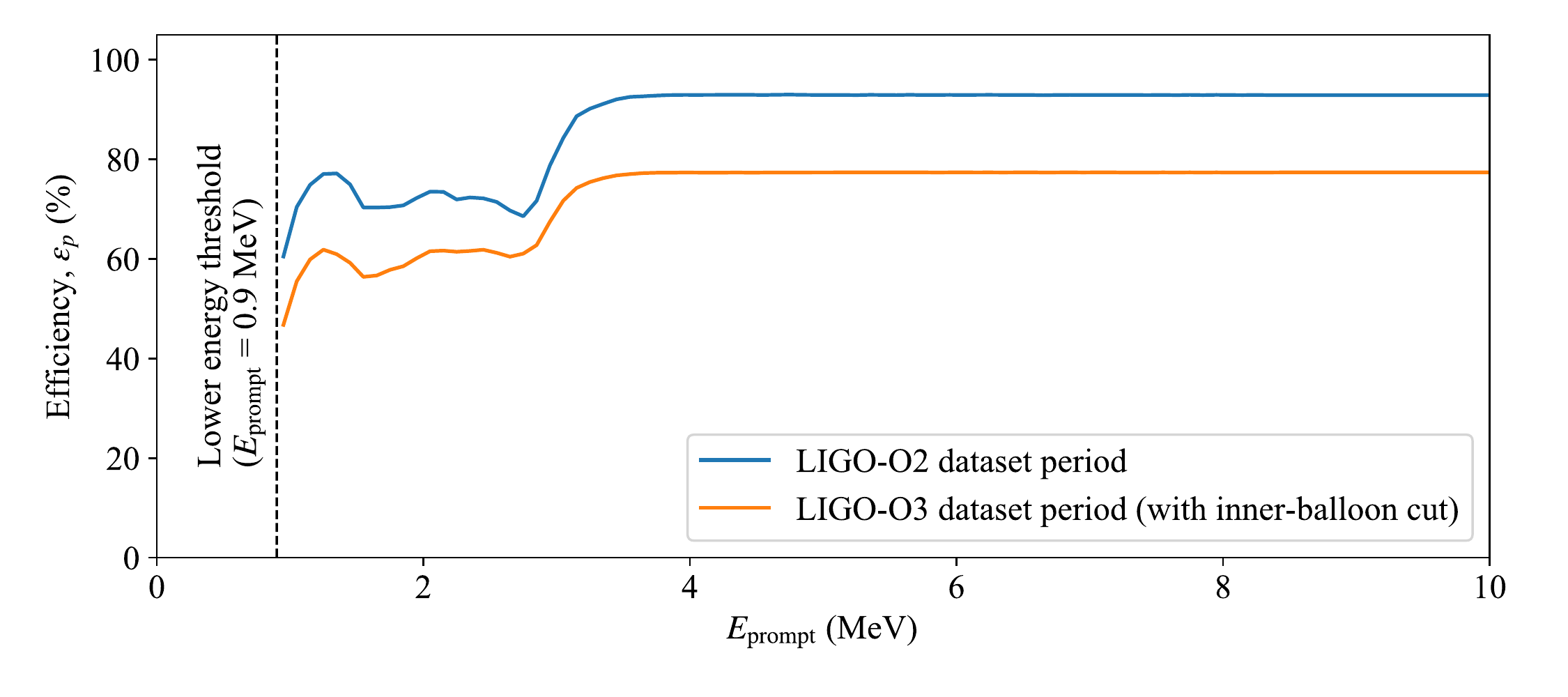}
    \caption{The electron antineutrino selection efficiencies as a function of the prompt energy. 
    The analysis period is divided into two datasets: the LIGO-O2 period in which we use of the full fiducial volume of the KamLAND detector (blue) and the LIGO-O3 period which includes the additional inner balloon cut described in the text.
    At a few MeV, the selection efficiencies are reduced by the likelihood selection to suppress the contamination of accidental coincidence. The vertical dashed line represents a lower energy threshold of $E_{\rm prompt} \geqq$ 0.9\,MeV.}
    \label{fig:AntivEff}
\end{figure}

The selection efficiencies are shown in Figure~\ref{fig:AntivEff}, as a function of the reconstructed prompt energy $(\epsilon_{\rm s}(E_{\rm prompt}))$.
The structure of the efficiency suppression around $E_{\rm prompt} \simeq 2\,{\rm MeV}$ is primarily derived from the accidental background spectrum shape. 
Above $E_{\rm{prompt}}=5.0$\,MeV,
at which point the accidental background contamination becomes negligibly small, the selection efficiencies in each dataset converge to 92.9\% and 77.4\%.

The dominant neutrino sources below 8\,MeV are the Japanese reactor power plants and geo-chemical radioactive decays in the Earth. 
After the Great East Japan Earthquake on 2011 March 11, most of the reactors in Japan were shut down and only a few have since been brought back online.
Other backgrounds are DC pairs of accidental radioisotopes, spallation products $^9$Li, and $^{13}$C$(\alpha,\ n)^{16}$O reaction.
Above $\sim$10\,MeV, fast neutrons from cosmic-ray muons and atmospheric neutrino interactions are the dominant contribution to the background~\citep{Gando_2012}.

%%%%%%%%%%%%%%%%%%%%%%%%%%%%%%%%%%%%%%%%%%%%%%%%%%%%%%%
\section{Coincidence event search} \label{sec:candidates}
This analysis is performed using a coincident time window of $\pm$500\,s around each of the 60 GW events listed in Table~\ref{tab:event1} and Table~\ref{tab:event2}. 
%our website\footnote[1]{\url{https://www.awa.tohoku.ac.jp/KamLAND/GW/2020/}}. 
The selected timing window is based on the largest expected time gap between GW events and neutrino events~\citep{BARET20111}.
This is sufficiently large to cover possible early neutrino emission scenarios as well as the neutrino time-of-flight delay from GW170729, the most distant GW source in this analysis.
For example, assuming the sum of neutrino mass limits and cosmological constants from \citet{aghanim2018planck}, and the neutrino mass-squared splittings from \citet{esteban2019global}, 
a neutrino with an energy of 1.8\,MeV, upper mass state of 60\,meV, traveling a distance of 2840\,Mpc will be delayed by approximately 86\,s relative to the GW.

The expected number of uncorrelated background events per $\pm$500\,s time window are estimated using off-time windows from the GW and found to be $4.08\times10^{-3}$ and $4.27\times10^{-3}$ for the KamLAND periods corresponding to LIGO-O2 and LIGO-O3, respectively.

\begin{figure}[htbp]
    \centering
    \includegraphics[width=1.0\linewidth]{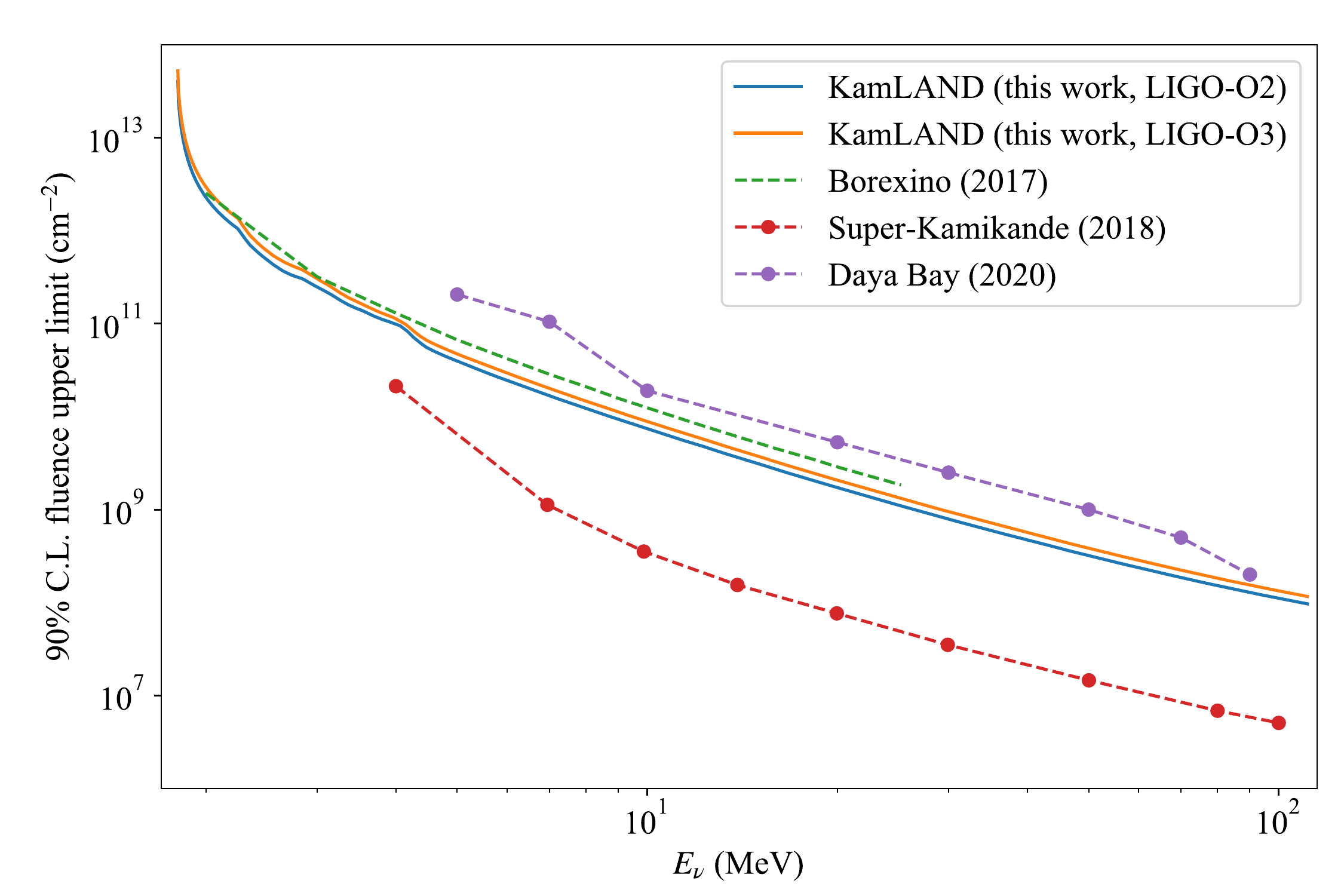}
    \caption{The 90\% C.L. electron antineutrino fluence upper limits for each GW. 
    The limits corresponding to events from LIGO-O2 are shown in blue, and events from LIGO-O3 are shown in orange.
    The difference between the two upper limits are primarily driven by the different selection efficiencies shown in Figure~\ref{fig:AntivEff}. 
    For comparison, the 90\% C.L. fluence upper limits on electron antineutrinos are also shown for Super-Kamiokande~\citep{Abe_2018}: GW170817; Borexino~\citep{Agostini_2017}: GW150914, GW151226, and GW170104; and Daya Bay~\citep{an2020search}: average of GW150914, GW151012, GW151226, GW170104, GW170608, GW170814, and GW170817.
    Borexino result as the un-binned analysis is shown as a green dashed line, Super-Kamiokande and Daya Bay results with binned analysis are shown as red dots and purple dots, respectively.
    }
    \label{fig:fluence}
\end{figure}

No IBD electron antineutrino events were found in the KamLAND dataset within $\pm$500\,s of each GW event.
Using the uncorrelated accidental background rates and zero observed signal events, the Feldman-Cousins method~\citep{PhysRevD.57.3873} is used to derive the 90\% confidence level (C.L.) upper limit on the number of detected electron antineutrinos.  
This is found to be $N_{90} = 2.435$ for each GW event in the LIGO-O2 and $N_{90} = 2.435$ for each GW event in the LIGO-O3. 
The upper limit ($F_{90}$) can then be used to place constraints on the neutrino fluence, as follows:
\begin{equation} \label{eq:fluence}
    F_{90} = \frac{N_{90}}{N_T\,\epsilon_{\rm live}\,\int \epsilon_{\rm s}(E_{\rm prompt} (E'_{\nu}))\,\sigma(E'_{\nu})\,\lambda(E'_{\nu})\,dE'_{\nu}},
\end{equation}
where $\sigma(E_{\nu})$ is the IBD cross section~\citep{STRUMIA200342}
and $\lambda(E_{\nu})$ is the neutrino energy spectrum.
In order to perform a model independent analysis from the neutrino emission mechanisms for various GW sources, we assume a monochromatic neutrino energy spectra. 
Hence, we calculate 90\% C.L. fluence upper limits on the electron antineutrinos for each GW event in LIGO-O2 and LIGO-O3 with 
\begin{equation}
    F_{90}(E_{\nu}) = \frac{N_{90}}{N_T\,\epsilon_{\rm live}\, \epsilon_{\rm s}(E_{\nu})\,\sigma(E_{\nu})},
\end{equation}
as shown in Figure~\ref{fig:fluence}.
The resulting upper limits on the electron antineutrino fluence are found to be between $10^{8}$--$10^{13}\,{\rm cm}^{-2}$.

We study the neutrino emission energy scales between two cases of GW sources for officially published and detail-known events during LIGO-O2: the BNS merger (BNS: GW170817), and six BBH mergers (BBHs: GW170104, GW170729, GW170809, GW170814, GW170818, GW170823). % as shown in Table~\ref{tab:event1}. 
Because of the $\pm$500\,s coincidence search timing window for each event, the total number of expected background events are $4.08\times10^{-3}$ for the BNS event and $2.45\times10^{-2}$ for the six BBH candidates.
Using the Feldman-Cousins method again with the 90\% C.L., for zero events observed, the upper limit on the number of neutrino events is $N_{90}^{\rm BNS} = 2.435$ and $N_{90}^{\rm BBHs} = 2.415$ for the BNS and the BBHs respectively.
According to \citet{PhysRevD.97.103001}, the assumption that the neutrino energy spectrum has a Fermi-Dirac distribution is reasonable for exploring the mechanism of neutrino emissions from the GW sources. 
Assuming the Fermi-Dirac distribution, the neutrino energy spectra can be written as
\begin{equation}
    \lambda_{\rm FD}(E_{\nu})  = \frac{1}{T^3 f_{2} (\eta)}\frac{E_{\nu}^2}{e^{E_{\nu}/T-\eta}+1}, 
\end{equation}
\begin{equation}
 f_{n}(\eta) = \int^{\infty}_{0} \frac{x^n}{e^{x-\eta}+1}dx,
\end{equation}
where we assume zero chemical potential and pinching factor $\eta=0$, the temperature is given as $T = \langle E \rangle /3.15$, and the average neutrino energy $\langle E \rangle = 12.7$\,MeV~\citep{PhysRevD.93.123015}. 
Integrating between the true electron antineutrino energy limits, $E_{\nu}$ = 1.8--111\,MeV, following Equation~(\ref{eq:fluence})
and assuming equal contribution from six neutrino species,
we obtain upper limits on the total fluence ($\mathcal{F}_{90}^{\rm BNS,\ BBH}$) in the Fermi-Dirac distribution case with 90\% C.L. as 
\begin{equation}
    \mathcal{F}_{90}^{\rm BNS} \le 2.04\times10^{10}\,{\rm cm}^{-2} %% all species
\end{equation}
for the BNS and
\begin{equation}
    \mathcal{F}_{90}^{\rm BBH} \le 2.02\times10^{10}\,{\rm cm}^{-2}
\end{equation}
for the BBH.
Considering the luminosity distances from the GW source,
we convert the total fluence ($\mathcal{F}_{90}$) to the total energy ($\mathcal{L}_{90}$) radiated in neutrinos from single source as
\begin{equation}
    \mathcal{L}_{90}^{\rm BNS,\ BBH} = \frac{\mathcal{F}_{90}^{\rm BNS,\ BBH}}{1/(4\pi D_{\rm eff}^2 \langle E \rangle)},
\end{equation}
where $D_{\rm eff}$ is the effective distance defined as $1/D_{\rm eff}^2 \equiv \sum_{i} 1/ D_i^2$ for every $i$-th GW events, and the central values are used to $D_i$.
Hence, the upper limits on the total energy are obtained as 
\begin{equation}
    \mathcal{L}_{90}^{\rm BNS} \leq 7.92\times10^{58}\,{\rm erg}
\end{equation}
based on the 40\,Mpc distance to the BNS event, and 
\begin{eqnarray}
    \mathcal{L}_{90}^{\rm BBH} \leq 8.22\times10^{60}\,{\rm erg}
\end{eqnarray}
for the BBHs based on the effective distance of 407.6\,Mpc, without accounting for neutrino oscillation effects.
The observed upper limits are found to be larger than the typical total energy 
radiated from supernovae $\mathcal{O}(10^{53})$\,erg~\citep{RevModPhys.62.801}.

%%%%%%%%%%%%%%%%%%%%%%%%%%%%%%%%%%%%%%%%%%%%%%%%%%%%%%%
\section{Summary} \label{sec:conclusion}
This paper searched for coincident IBD electron antineutrinos in KamLAND with the 60 GW events associated with the second and third observing runs of the LIGO detector. 
No coincident signal was observed within a $\pm$500\,s timing window around each GW event. 
The 90\% C.L. electron antineutrino fluence upper limit for each GW, assuming a mono-energetic neutrino flux, was presented for neutrino energies between 1.8\,MeV and 111\,MeV.
We set the most strict upper limit on each GW event in the LIGO-O2 dataset below 3.5\,MeV neutrino energies.
For the LIGO-O3 dataset, this is the first result of an MeV-scale energy coincidence neutrino search.

The obtained upper limits on the total energy radiated from GW source class, BNS or BBH, in LIGO-O2 with the assumption of a Fermi-Dirac neutrino energy distribution, are found to be $7.92\times10^{58}\,{\rm erg}$ and $8.22\times10^{60}\,{\rm erg}$, respectively.
These results depend on the number of GW events and distances. 
This limit will be improved once the candidate events of LIGO-O3 are published.

In the future, the mechanism of neutrino emission may be constrained and explored by combining with multi-messenger astronomy: GeV/TeV neutrino detectors, X-ray/gamma-ray telescopes, and gravitational wave detectors.
The KamLAND detector continues to take physics data while running in the KamLAND-Zen~800 configuration and is monitoring for transient astrophysical events. 
The recently implemented online monitor at KamLAND~\citep{Asakura_2016} also readily searches for correlations with transient events and reports the results to the Gamma-ray Coordinates Network (GCN) and/or the Astronomer's Telegram (ATel).

%%%%%%%%%%%%%%%%%%%%%%%%%%%%%%%%%%%%%%%%%%%%%%%%%%%%%%%
\acknowledgments
The KamLAND experiment is supported by 
JSPS KAKENHI Grants 
19H05803; % 2019-06-28 \UTF{2013} 2024-03-31, Inoue, Shingakujutsu-ChikaUtyu (https://kaken.nii.ac.jp/ja/grant/KAKENHI-PLANNED-19H05803/)
the World Premier International Research Center Initiative (WPI Initiative), MEXT, Japan; 
Netherlands Organisation for Scientific Research (NWO); %Patric
and under the U.S. Department of Energy (DOE) Contract 
No.~DE-AC02-05CH11231,
the National Science Foundation (NSF) No.~NSF-1806440, %Lindley
NSF-2012964, %Chris
as well as other DOE and NSF grants to individual institutions.  
The Kamioka Mining and Smelting Company has provided service for activities in the mine.  
We acknowledge the support of NII for SINET4. 
This work is partly supported by 
the Graduate Program on Physics for the Universe (GP-PU), 
and the Frontier Research Institute for Interdisciplinary Sciences, Tohoku University.

%%%%%%%%%%%%%%%%%%%%%%%%%%%%%%%%%%%%%%%%%%%%%%%%%%%%%%%
\bibliographystyle{aasjournal}
\bibliography{main}{}

\end{document}